\documentclass[prl,showpacs,preprintnumbers,amsmath,amssymb,superscriptaddress]{revtex4}
\usepackage{graphicx}
\usepackage{dcolumn}
\usepackage{bm}
\begin{document}

\title{Partial and Quasi Dynamical Symmetries in Nuclei }
  

\author{P.~Van~Isacker}
\affiliation{Grand Acc\'el\'erateur d'Ions Lourds, CEA/DSM - CNRS/IN2P3\\
Bd Henri Becquerel, BP 55027, F-14076 Caen Cedex 5, France}
\begin{abstract}
One of the interesting aspects in the study of atomic nuclei
is the strikingly regular behaviour many display
in spite of being complex quantum-mechanical systems,
prompting the universal question of how regularity emerges out of complexity.
It is often conjectured that symmetries play a pivotal role
in our understanding of this emerging simplicity.
But most symmetries are likely to be broken, partial or both.
Under such more realistic conditions,
does the concept of symmetry still provide a basis for our understanding of regularity?
I suggest that this requires the enlarged concepts of partial and quasi dynamical symmetry.
\end{abstract}

\maketitle 

\section{Dynamical Symmetry}
Symmetries are of central importance to physics.
Classically they determine constants of motion and conserved quantities.
In quantum mechanics they provide quantum numbers
to label eigenstates, determine degeneracies
and lead to selection rules that govern transition processes.
A derived notion is that of dynamical symmetry
concerning which, unfortunately, considerable semantic confusion exists
and which is defined here following Iachello~\cite{Iachello06}.
While a hamiltonian with a given symmetry
commutes with all transformations associated with that symmetry,
this is no longer the case for a hamiltonian with a dynamical symmetry
which only commutes with certain combinations of those transformations (called Casimir operators)
which commute among themselves.
As a result, eigenstates of a hamiltonian with a dynamical symmetry,
as compared to those of one with a symmetry,
are characterized by the same labels
and obey the same selection rules
but the feature of spectral degeneracy is relaxed.
The symmetry breaking occurs for the hamiltonian but not for its eigenstates.

While dynamical symmetry might be unfamiliar as a term,
the concept is used extensively in diverse areas of physics,
and in particular in nuclear physics.
Notable examples are Racah's pairing model~\cite{Racah43}
and Elliott's rotation model~\cite{Elliott58},
and their many extensions that can be formulated
in terms of dynamical symmetries~\cite{Frank09}.
Lesser known are two further generalizations
which have been developed over the last two decades,
namely partial dynamical symmetry (PDS)~\cite{Leviatan11}
and quasi dynamical symmetry (QDS)~\cite{Rochford88},
and which are the subject matter of the present article.
I will illustrate the basic ideas behind these extensions
in the context of a simplified version of a nuclear model,
while emphasizing that they are of generic applicability to any many-body problem
as long as it can be formulated in an algebraic language.

\section{The Interacting Boson Model as a Symmetry Laboratory}
The interacting boson model (IBM) proposed by Arima and Iachello~\cite{Arima75}
is ideally suited to illustrate the notion of dynamical symmetry
and its above-mentioned generalizations.
The model in its simplest version assumes a system of $N$ interacting bosons
that have either spin 0 ($s$) or spin 2 ($d$).
The bosons can be thought of as pairs of nucleons (neutrons or protons)
in the valence shell of the nucleus~\cite{Iachello87}.
Alternatively, the $d$ bosons can be associated
with quadrupole oscillations of the nuclear surface~\cite{Ginocchio80}.
Over the years the IBM has revealed a rich algebraic structure,
allowing a simple yet detailed description of many collective properties of nuclei.
Numerical techniques exist to solve the secular equation associated with a general IBM hamiltonian.
An additional characteristic feature of the model is that,
for particular choices of the boson energies and the boson-boson interactions,
this equation admits analytic solutions.
This happens when the hamiltonian can be written
as a combination of commuting Casimir operators of U(6),
the algebra formed by the $s$ and $d$ bosons, and its subalgebras.
The analysis of Arima and Iachello shows~\cite{Arima76}
that three such cases (that is, three dynamical symmetries) exist,
corresponding to nuclear spectra with a vibrational [U(5)], a rotational [SU(3)]
or a  $\gamma$-soft [SO(6)] character.
The essential symmetry properties of the IBM
are captured by the hamiltonian of the so-called extended consistent-Q formalism (ECQF)~\cite{Warner83},
\begin{equation}
\hat H_{\rm ECQF}=
\omega\left[
(1-\xi)\hat n_d-\frac{\xi}{4N}\hat Q^\chi\cdot\hat Q^\chi
\right],
\label{e_ecqf}
\end{equation}
where $\hat n_d$ is an operator that counts the number of $d$ bosons
and $\hat Q^\chi$ is the quadrupole operator of the model containing a parameter $\chi$.
The structural properties of the ECQF depend on the two parameters $\xi$ and $\chi$
since $\omega$ is but an overall scale parameter.
The conventional ranges of the two structural parameters
are $0\leq\xi\leq1$ and $-\sqrt{7}/2\leq\chi\leq0$,
and their extreme values $(\xi,\chi)=(0,{\rm anything})$, $(1,-\sqrt{7}/2)$ and $(1,0)$ 
correspond to the U(5), SU(3) and SO(6) dynamical symmetries (or `limits' for short), respectively.
The entire physical parameter space
can therefore be represented on a so-called Casten triangle~\cite{Casten81},
each point of which corresponds to a specific choice of the parameters in the ECQF hamiltonian.
(It is possible to extend this triangle to include the region $0\leq\chi\leq+\sqrt{7}/2$;
this extension is important
as regards the quantum phase transitions of the ECQF hamiltonian~\cite{Jolie01}
but is not considered here in the discussion of its symmetry properties.)

One way to determine the symmetry properties of a given hamiltonian
is with the help of its `wave-function entropy' (WFE)~\cite{Cejnar98}.
For one of its eigenstates this quantity is defined as $-\sum_i\alpha_i^2\ln\alpha_i^2$
where $\alpha_i$  are the stateÕs amplitudes in a given basis
which in the IBM are usually taken to be U(5), SU(3) or SO(6).
A vanishing WFE therefore implies an exact U(5), SU(3) or SO(6) dynamical symmetry.
For example, any eigenstate of an SU(3) hamiltonian has zero WFE in the SU(3) basis
(since one amplitude $\alpha_i$ equals 1 and all others 0)
but a large one in U(5).
Figure 1 shows the averaged WFE for all eigenstates of the ECQF hamiltonian
with angular momentum $L=0$ in the three different bases.
As expected, a blue region (low WFE) is found
around the vertex that corresponds to the basis used.
It is tempting to conclude from this figure
that the interior region of the triangle
is not amenable to any symmetry treatment.
That conclusion would be wrong.
\begin{figure}
$\begin{array}{ccc}
\includegraphics[width=5cm]{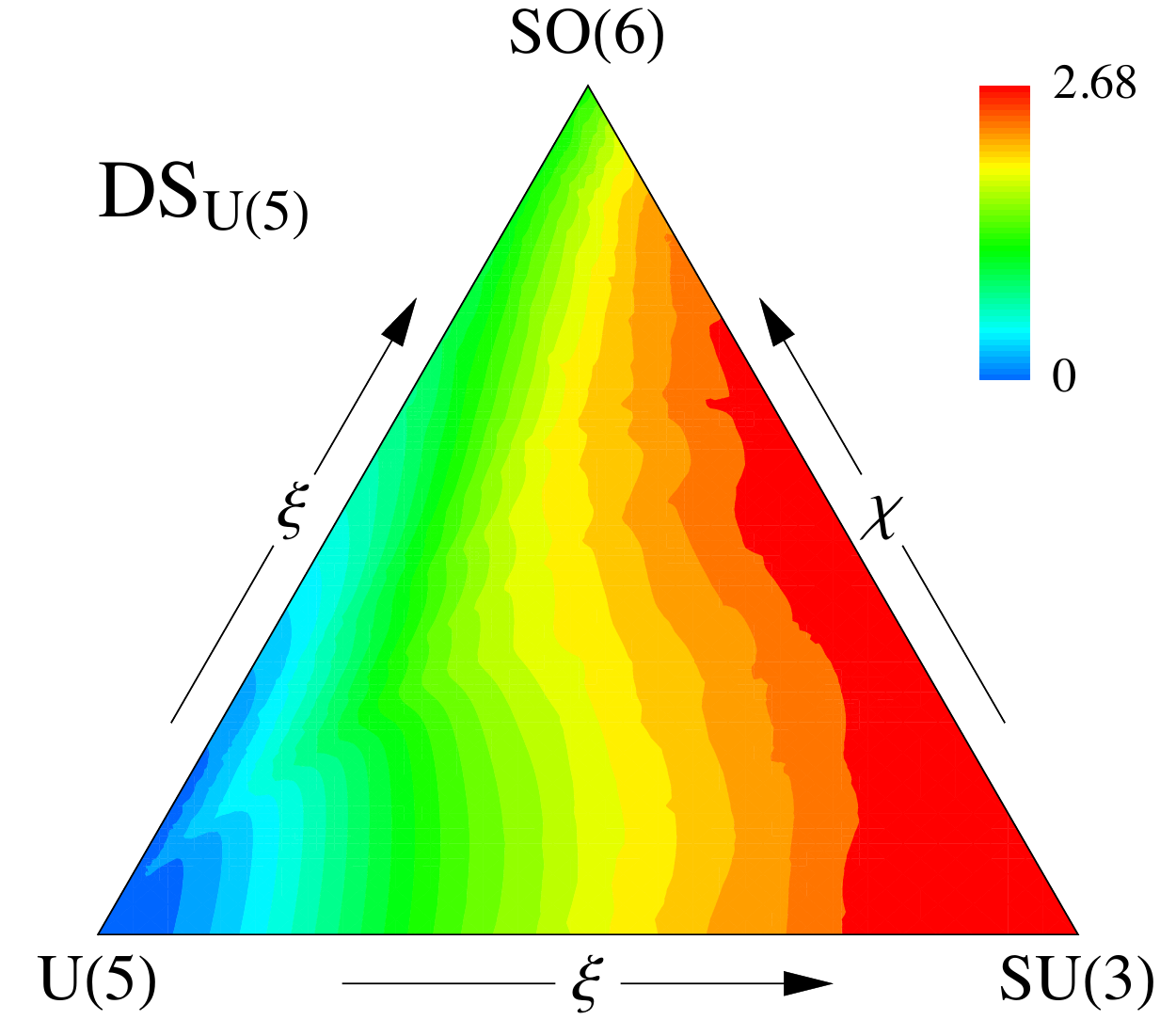}&
\includegraphics[width=5cm]{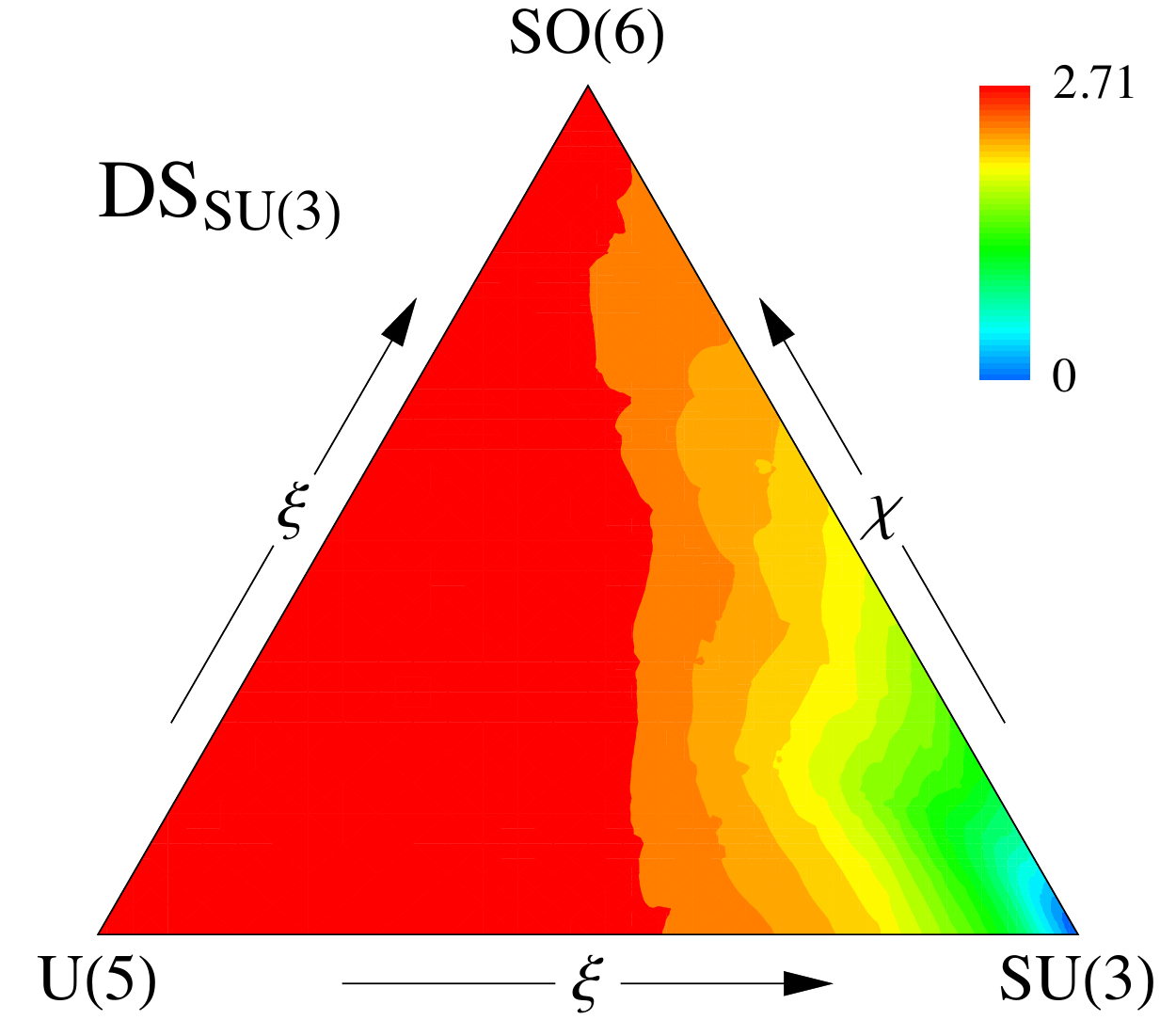}&
\includegraphics[width=5cm]{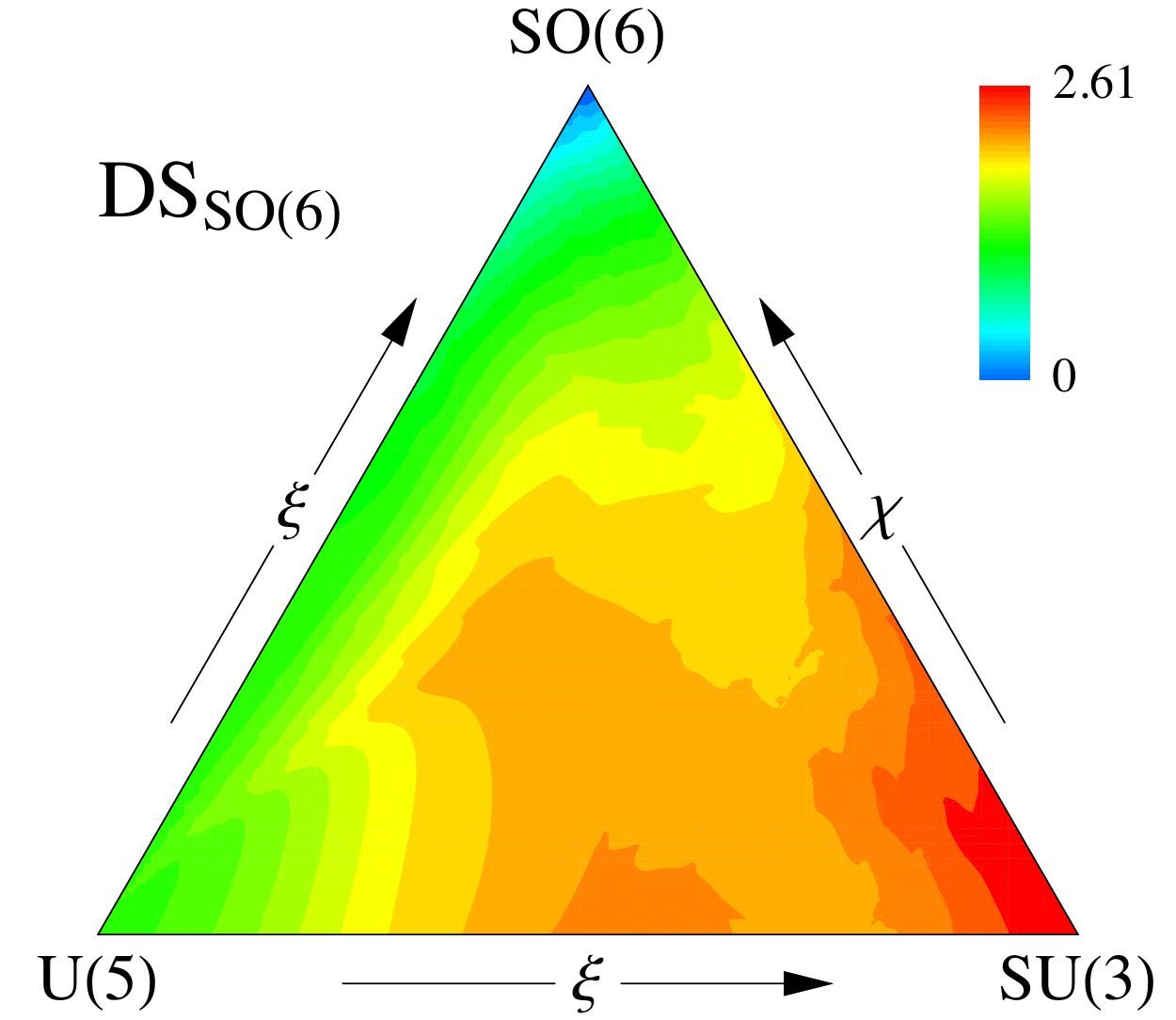}
\end{array}$
\caption{
Illustration of the concept of dynamical symmetry (DS) for the ECQF hamiltonian.
The plots show the wave-function entropy,
in the three different bases U(5), SU(3) and SO(6) (left, middle and right),
averaged for all eigenstates with angular momentum $L=0$ and boson number $N=15$.
This quantity is a measure of dynamical symmetry 
and vanishes when all quantum numbers of the basis are exactly conserved for all eigenstates.}
\end{figure}

The results of figure 1 are found
if one considers the WFE of all eigenstates (with $L=0$)
with respect to all quantum numbers in a given basis.
But in most applications one's interest is confined to a few low-energy eigenstates.
In addition, interesting symmetry properties may arise
if some symmetries are broken while others are conserved.
To uncover the existence of such cases,
one considers the WFE of a subset of eigenstates of the ECQF hamiltonian
and/or the decomposition of these eigenstates onto subspaces
characterized by a single label of a particular subalgebra.
There are many combinations of this kind
and figure 2 shows the WFE for three choices of eigenstates and/or labels in an SO(6) basis
(again for $L=0$ but similar results are obtained for other $L$).
From left to right the figure illustrates the three types of partialities
as defined in Ref.~\cite{Leviatan11}:
(PDS1) some eigenstates carry all the labels proper to a particular dynamical symmetry,
(PDS2) all eigenstates carry some of its labels
and (PDS3) some eigenstates carry some labels.
A surprising result of figure 2, only recently recognized~\cite{Kremer},
is shown in the plot PDS3$_{\rm SO(6)}$
which quantifies the conservation of the SO(6) label in the ground state
and uncovers an entire band of ECQF hamiltonians
with approximate ground-state SO(6) symmetry. 
\begin{figure}
$\begin{array}{ccc}
\includegraphics[width=5cm]{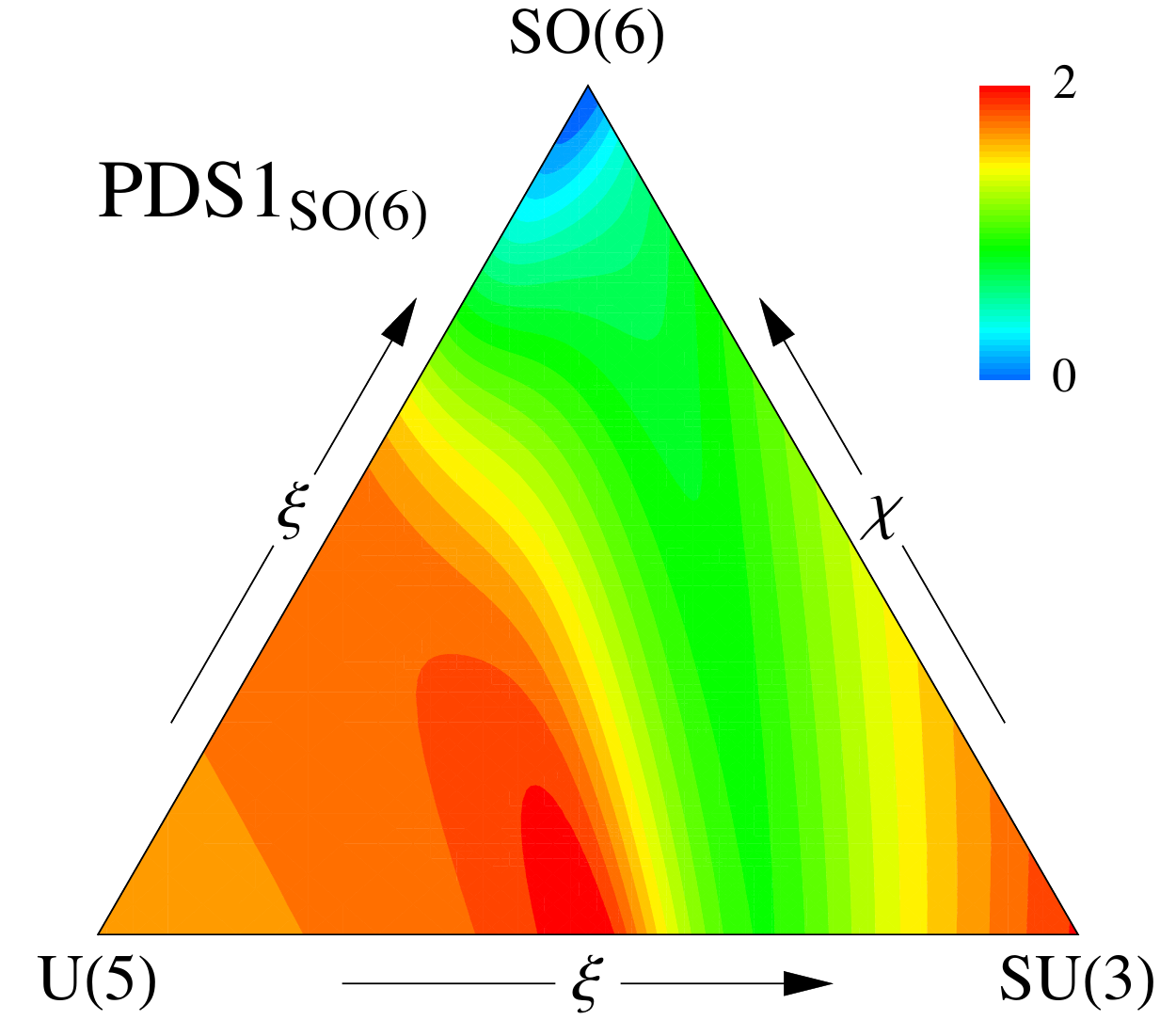}&
\includegraphics[width=5cm]{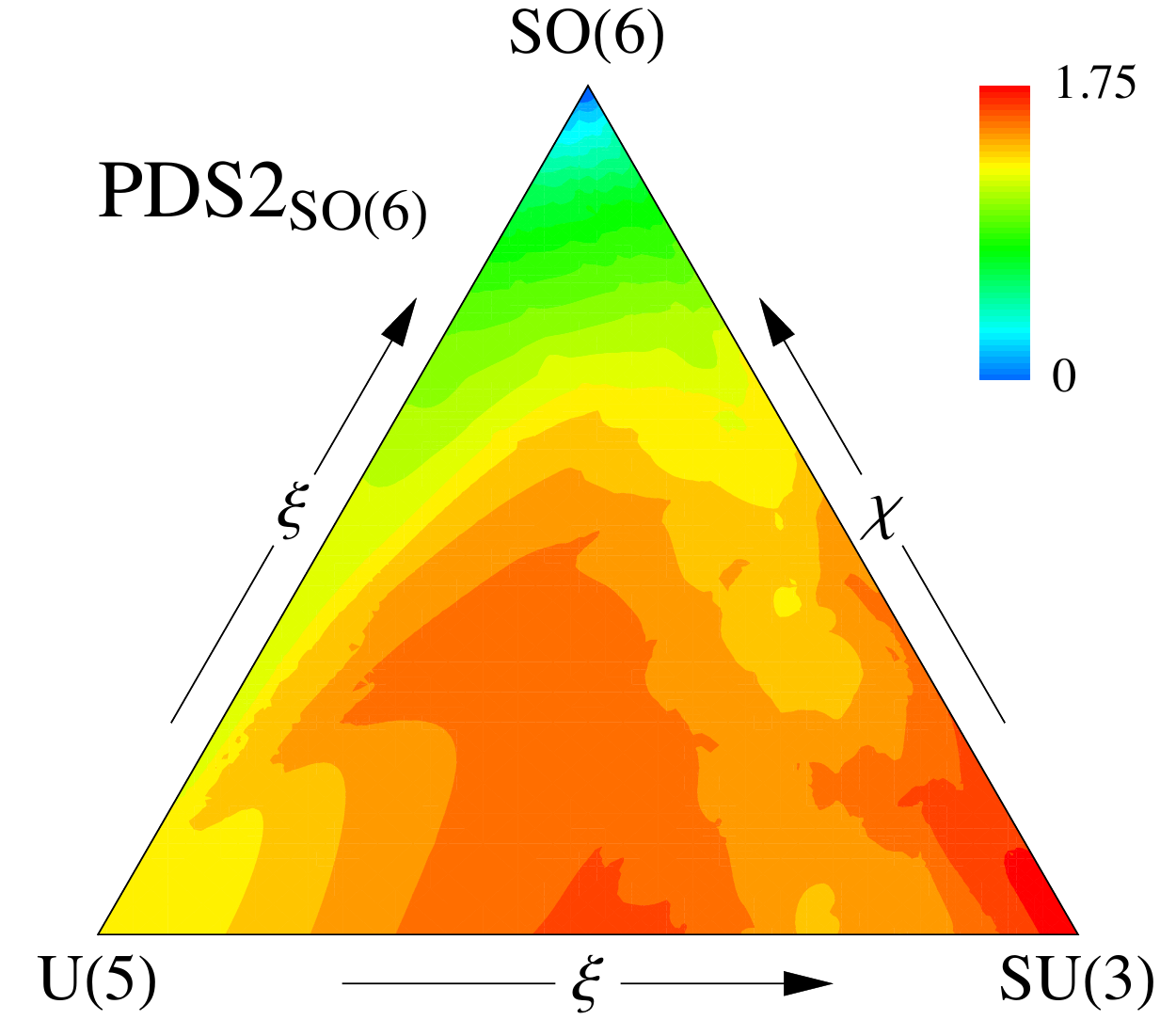}&
\includegraphics[width=5cm]{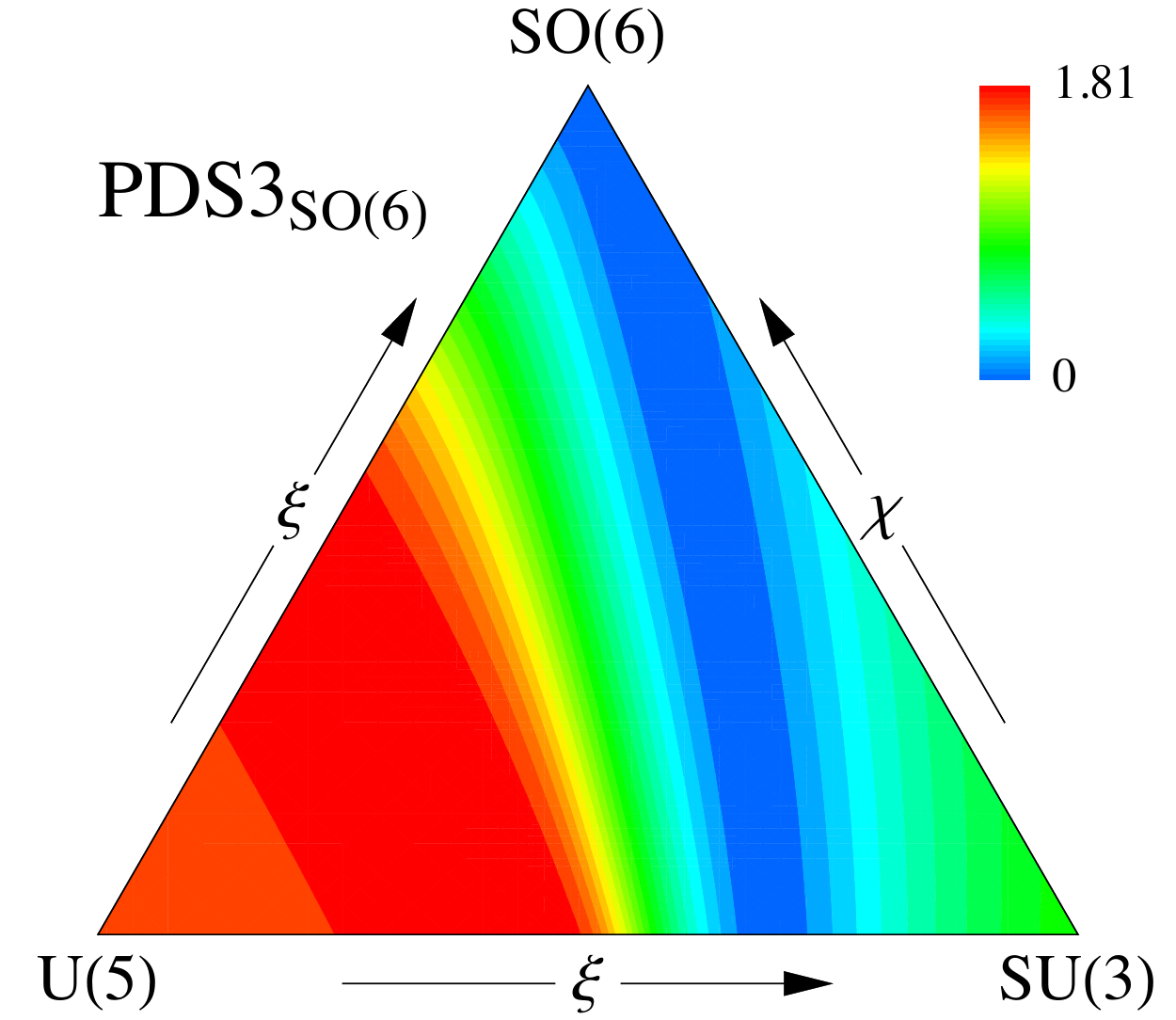}
\end{array}$
\caption{
Illustration of the concept of partial dynamical symmetry (PDS) for the ECQF hamiltonian.
The plots show the wave-function entropy (WFE), in the SO(6) basis,
of eigenstates with angular momentum $L=0$ and boson number $N=15$.
From left to right is shown:
(PDS1) WFE of the yrast eigenstate with respect to all labels of the SO(6) limit,
(PDS2) WFE of all eigenstates with respect to the SO(6) label
and (PDS3) WFE of the yrast eigenstate with respect to the SO(6) label.}
\end{figure}

Figure 3 provides another illustration
of the baffling multitude of symmetry features of the Casten triangle.
The plot PDS2$_{\rm SO(5)}$ indicates to what extent the label of the SO(5) subalgebra---associated
with $d$-boson seniority---is
conserved for all eigenstates of the ECQF hamiltonian,
and it turns out that the conservation of this label
is exact for the entire U(5)--SO(6) transitional hamiltonian.
This feature was noted a long time ago~\cite{Leviatan86}
and it can be shown, more generally,
that the U(5)--SO(6) transitional hamiltonian is in fact integrable~\cite{Pan98}.
Another noteworthy result is shown in the plot PDS3$_{\rm SO(5)}$ of figure 3
which proves that a large area in the Casten triangle
corresponds to ECQF hamiltonians with approximate SO(5) symmetry in the ground state.
One may therefore expect selection rules
associated with this symmetry to have a wide validity in nuclei.
\begin{figure}
$\begin{array}{ccc}
\includegraphics[width=5cm]{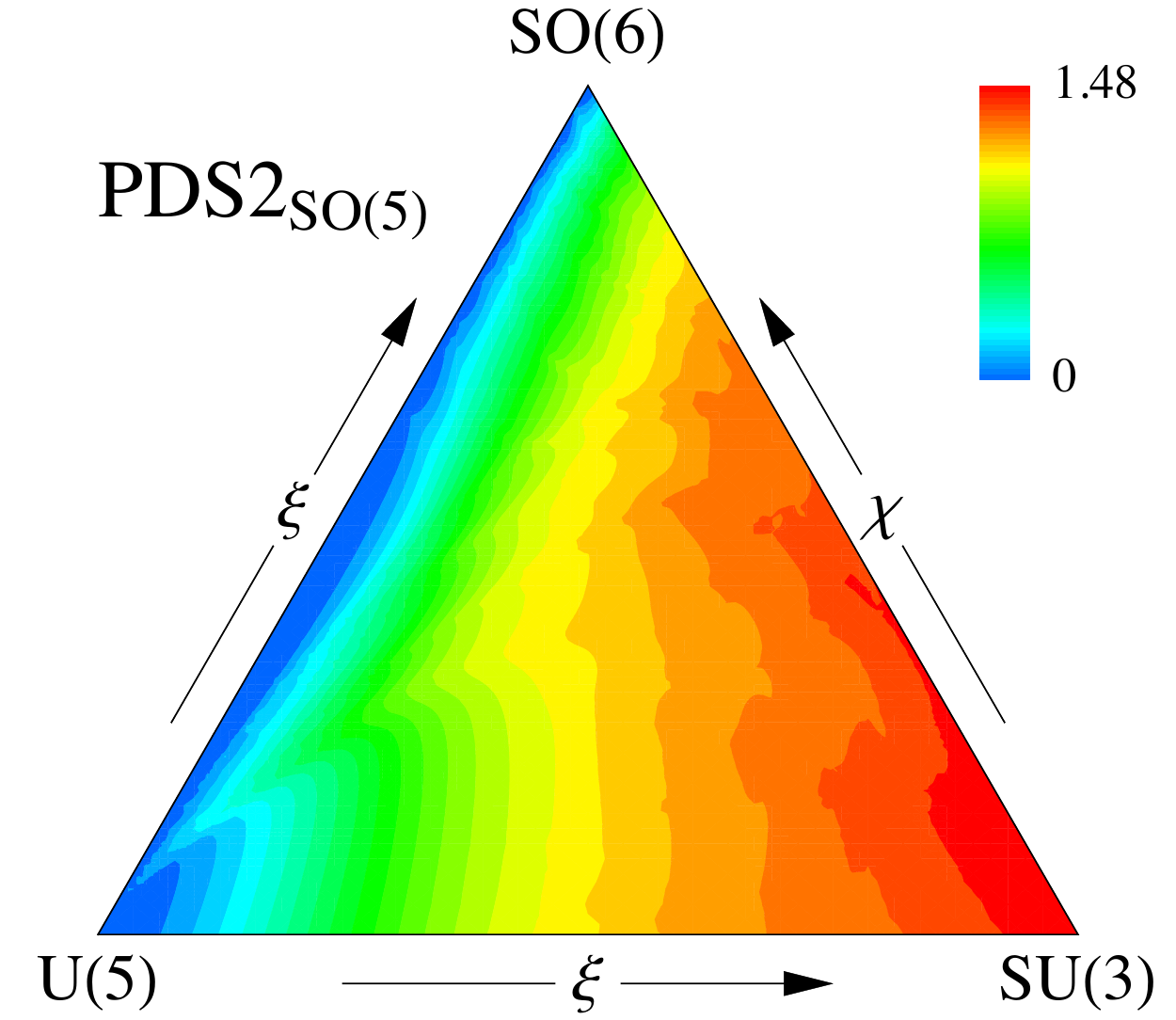}&
\includegraphics[width=5cm]{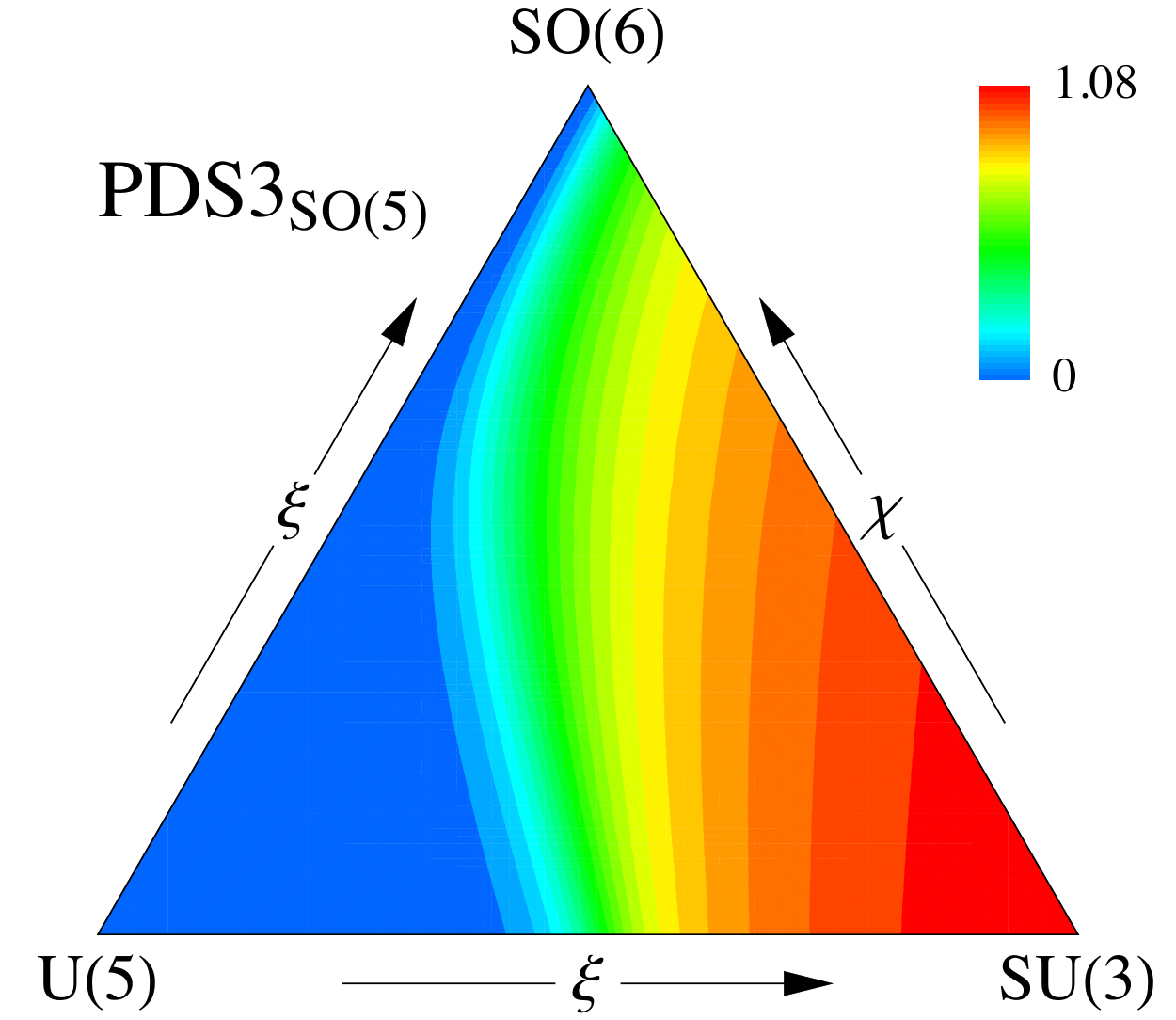}
\end{array}$
\caption{
Illustration of the concept of partial dynamical symmetry (PDS) for the ECQF hamiltonian.
The plots show the wave-function entropy (WFE), with respect to the SO(5) label,
of eigenstates with angular momentum $L=0$ and boson number $N=15$.
From left to right is shown:
(PDS2) WFE of all eigenstates with respect to the SO(5) label
and (PDS3) WFE of the yrast eigenstate with respect to the SO(5) label.}
\end{figure}

One can similarly investigate the persistence of quasi-dynamical symmetry (QDS) in the ECQF hamiltonian.
While QDS can be defined mathematically in terms of embedded representations~\cite{Rowe88},
its physical meaning is that several observables
associated with a particular subset of eigenstates
may be consistent with a certain symmetry,
which in fact is broken in the hamiltonian and its eigenstates.
This typically occurs for a hamiltonian
transitional between two dynamical symmetries
which retains, for a certain range of its parameters,
the characteristics of one of those limits~\cite{Rowe98}.
This `apparent' symmetry is due to a coherent mixing
of representations in selected eigenstates.
The criterion for the validity of QDS
is not the WFE of individual eigenstates in a certain basis
but rather the similarity of the decomposition
in that basis for a subset of eigenstates.
Figure 4 shows a quantitative measure of this similarity
for yrast eigenstates of the ECQF hamiltonian with $L=0,2,\dots,10$.
It is defined as $\sqrt{1-\bar\Theta}$  where $\bar\Theta$
is the average of $\Theta_{LL'}=\sum_i\alpha_i^L\alpha_i^{L'}$ for all pairs $L\neq L'$,
where $\alpha_i^L$ are the amplitudes of the yrast eigenstates
with angular momentum $L$ in a U(5), SU(3) or SO(6) basis.
It is seen that large areas of the Casten triangle are blue ({\it i.e.}, display QDS).
\begin{figure}
$\begin{array}{ccc}
\includegraphics[width=5cm]{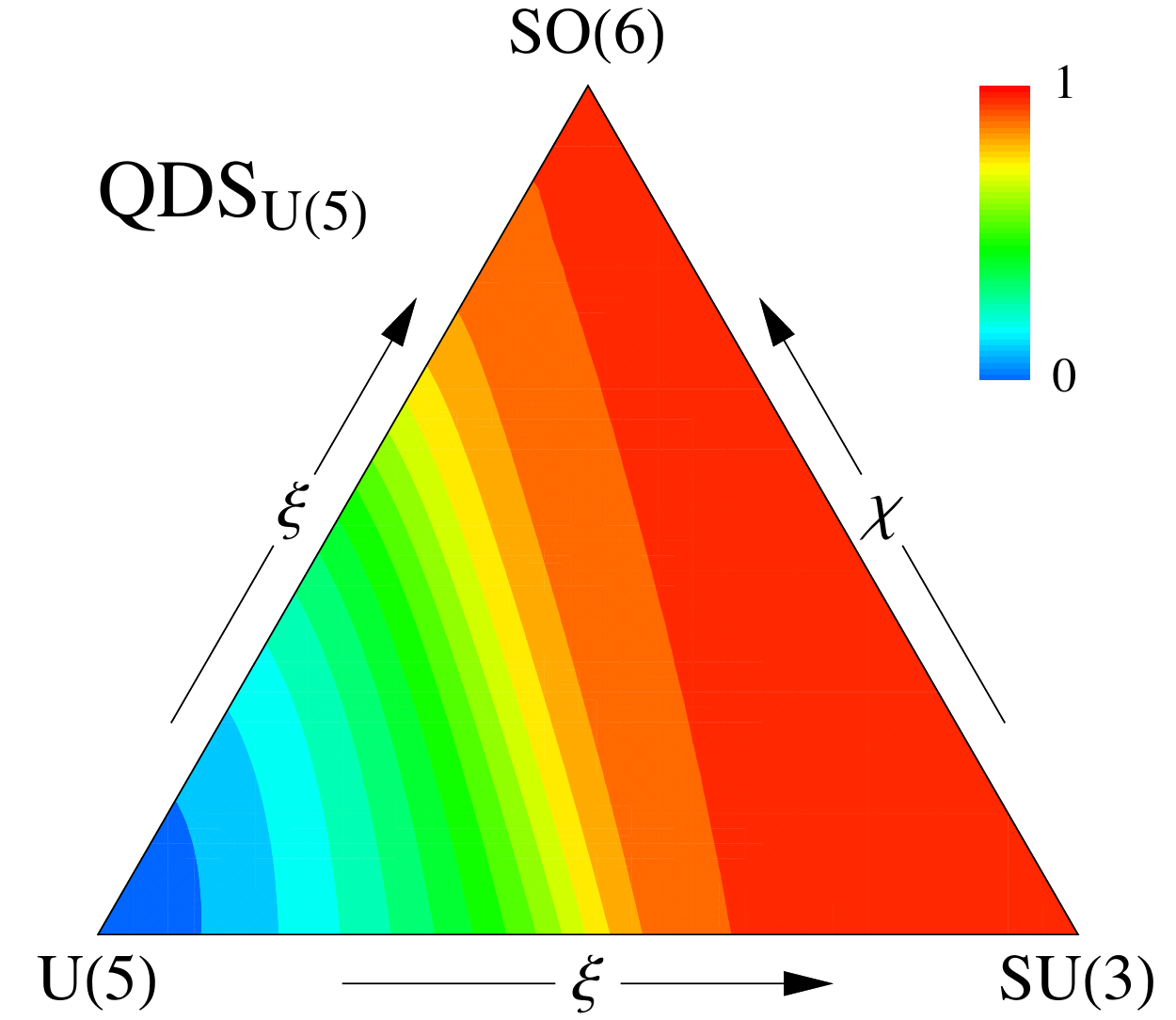}&
\includegraphics[width=5cm]{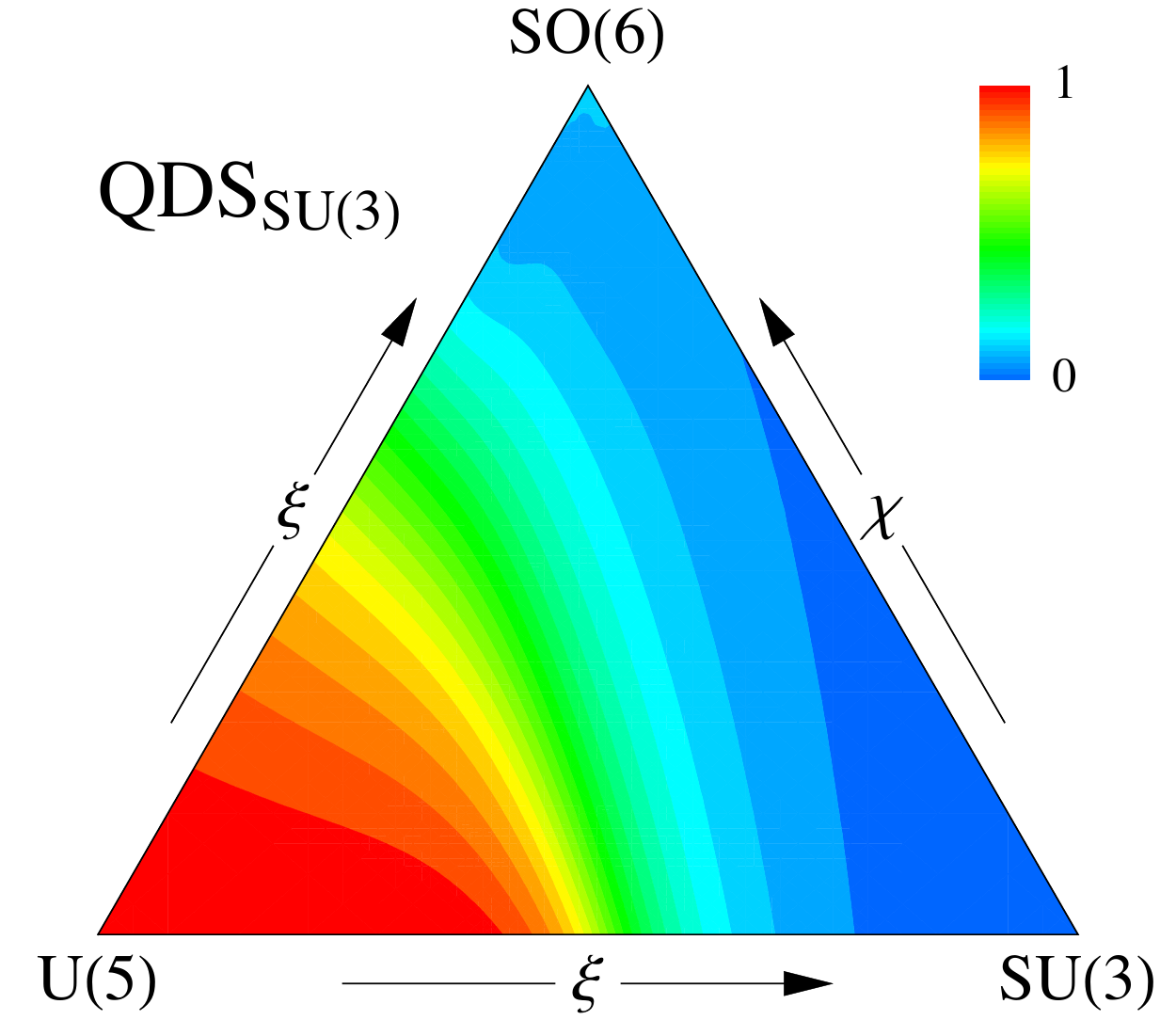}&
\includegraphics[width=5cm]{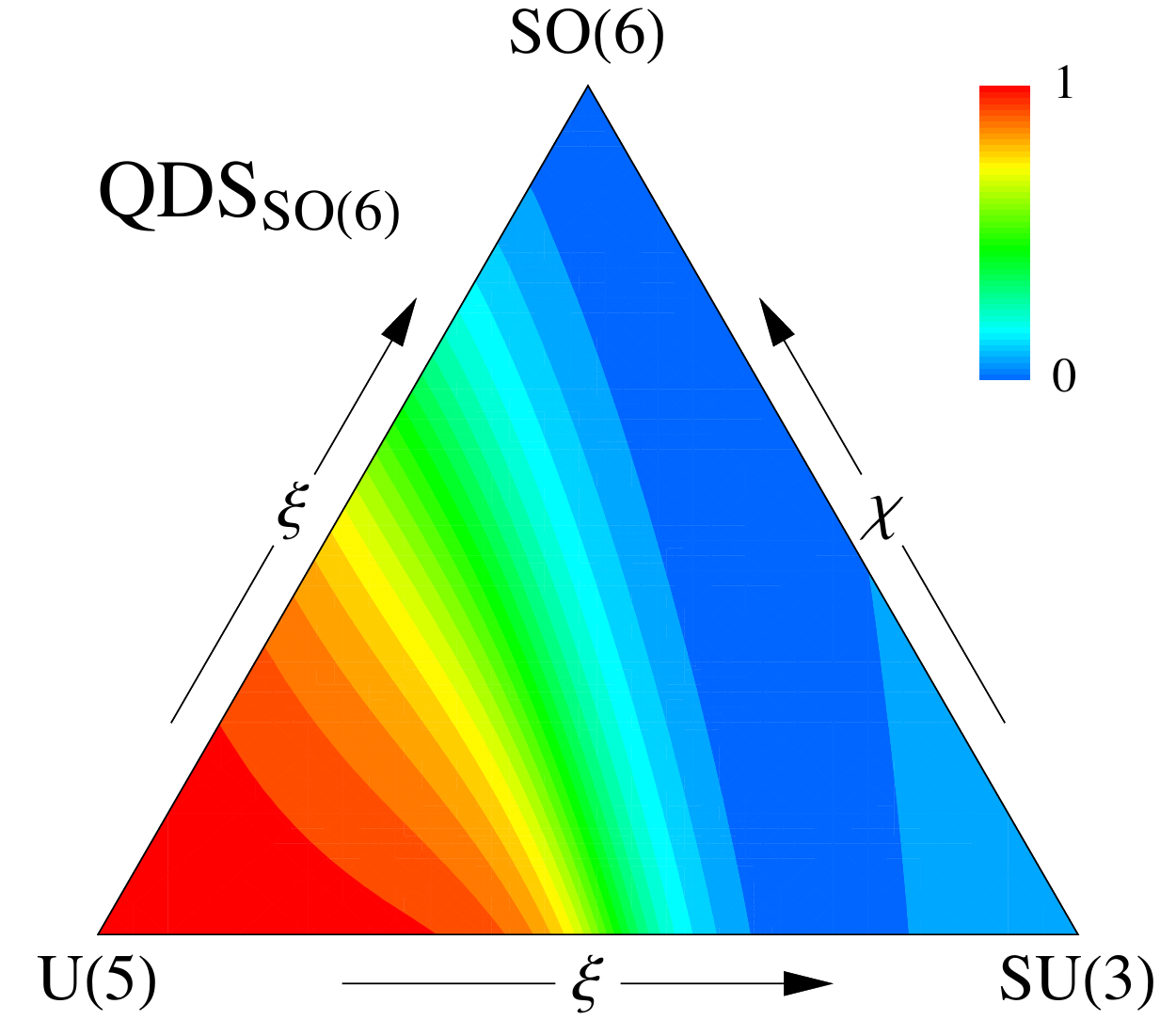}
\end{array}$
\caption{
Illustration of the concept of quasi dynamical symmetry (QDS)
for the yrast eigenstates of the ECQF hamiltonian
in the three different bases U(5), SU(3) and SO(6) (left, middle and right).
The plots show a quantitative measure (as defined in the text)
of the similarity of the decomposition in the different bases for the yrast eigenstates
with even angular momentum $L=0$ to 10 and boson number $N=15$.}
\end{figure}

There are obvious connections between the symmetry concepts of PDS and QDS,
as illustrated by comparing the plots PDS3$_{\rm SO(6)}$ in figure 2
and QDS3$_{\rm SO(6)}$ in figure 4 which have a similar structure.
Even more surprising is the fact that areas of the Casten triangle exist
where the partial conservation of one symmetry [SO(6) PDS]
is consistent with a coherent mixing of another, incompatible symmetry [SU(3) QDS].
The contrast of the results shown in figures 2 to 4 with those of figure 1 is startling.
While dynamical symmetries are restricted to small regions in the parameter space
(the blue areas in figure 1)
and therefore can be expected to have only marginal applicability in nuclei,
this is not the case for the extended concepts of PDS and QDS---see
figures 2 to 4 where large bands of blue abound in the triangles.
It is in fact difficult to find a spot in the Casten triangle
that is not amenable to some symmetry treatment!

The ECQF is used here only to illustrate with figures
the notions of PDS and QDS,
and in no way do these results represent the complete symmetry analysis of the IBM.
A general IBM hamiltonian with up to two-body interactions
allows the occurrence of exact dynamical symmetries of various partialities~\cite{Leviatan96},
some of which are not or only approximately present in the ECQF.
In addition, dynamical symmetries exist
corresponding to different phase conventions~\cite{Shirokov98},
and these should be taken into account in a full symmetry study.
Finally, three-body interactions between the bosons,
which are to be expected given their composite nature,
further enrich the symmetry features of the IBM~\cite{Isacker99}.
It is remarkable that close to forty years after the proposal by Arima and Iachello,
the full symmetry content of the IBM still remains to be uncovered.

\section{An Example from the Shell Model: Seniority Isomers}
Although most applications of PDS and QDS
have been considered in the context of the IBM,
they are by no means restricted to this model.
As an illustration of this statement, I present an example taken from the shell model.
The starting point is the original analysis of Racah~\cite{Racah43}
who proved that the pairing interaction between identical fermions in a $j$ shell
is characterized by the seniority quantum number (usually denoted by $\upsilon$)
which counts the number of nucleons
not in pairs coupled to angular momentum zero.
Subsequent studies showed that seniority has a much wider applicability
in the sense that any two-body interaction between identical nucleons
conserves this quantum number as long as the angular momentum $j$ of the shell
is smaller than or equal to 7/2~\cite{Shalit63}.

There are clear observational consequences
associated with the conservation of seniority,
one of them being vanishing E2 matrix elements between levels with the same seniority
when the $j$ shell is half filled~\cite{Shalit63}.
These findings are at the basis of the existence of so-called seniority isomers~\cite{Grawe97},
nuclear levels with a half-life typically in the micro-second range,
whose decay is hindered by this seniority selection rule.
In the left part of figure 5 is shown the spectrum of $^{94}$Ru~\cite{Mills07},
which has an $8^+$ isomer at 2.644~MeV with a half-life of 71~$\mu$s.
The isomeric character of this level is a consequence of the slow E2 decay
and the small energy difference with the $6^+$ level below it.
In a first approximation, $^{94}$Ru can be treated as a semi-magic nucleus
with four protons in the $1g_{9/2}$ shell to which a seniority classification can be applied.
The expected seniority spectrum is shown in the middle of figure 5
and it is seen that two $4^+$ and two $6^+$ levels occur close in energy,
with seniority $\upsilon=2$ (blue) and $\upsilon=4$ (red), respectively.
Since for a general interaction
seniority is not a conserved quantum number in a shell with $j=9/2$,
one would expect the states with the same angular momentum to mix,
which in turn would modify the E2 decay of the $8^+$ level.
This is not the case however, since it can be shown~\cite{Escuderos06}
that the states indicated in red have exact seniority $\upsilon=4$ for any two-body interaction.
This is an example of a PDS in a fermion system,
in this case associated with seniority.
As a result of PDS, no mixing occurs,
the predictions of the seniority scheme hold true
and the isomeric character of the $8^+$ level
depends essentially on its position relative to the $6^+$ level with seniority $\upsilon=4$.
\begin{figure}
\includegraphics[width=10cm]{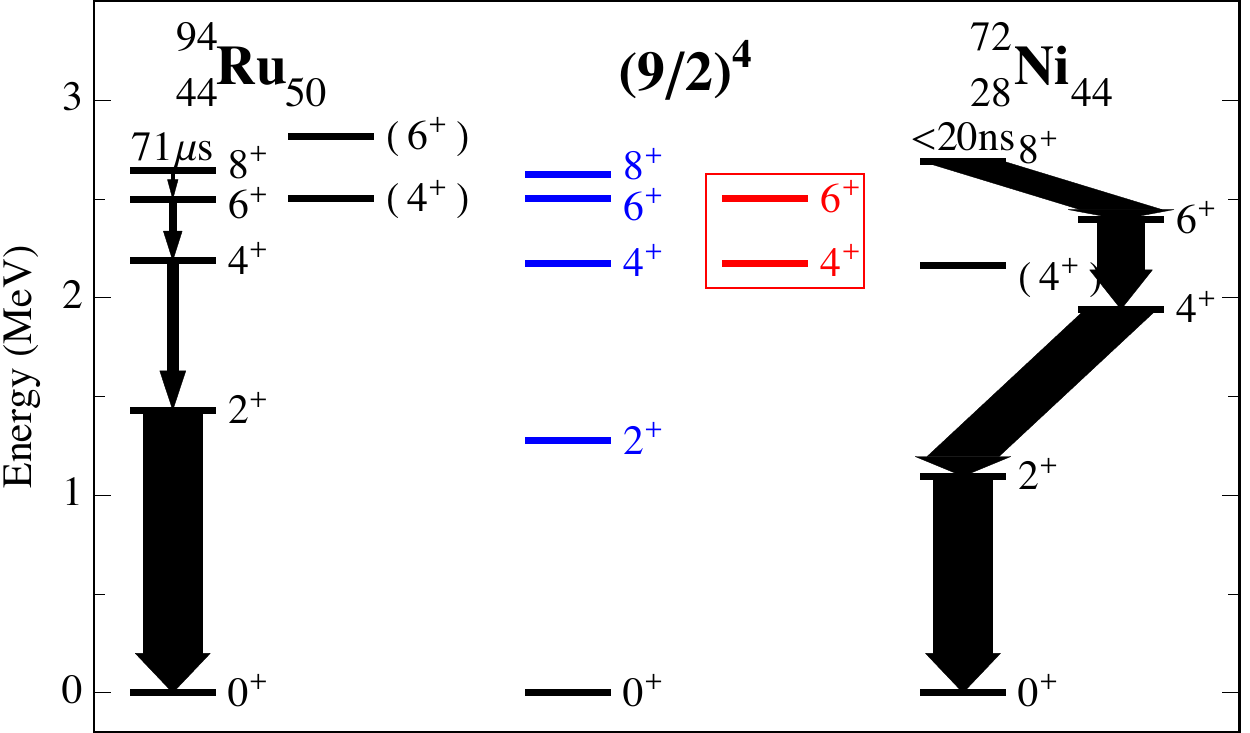}
\caption{
Energy spectra of $^{94}$Ru and $^{72}$Ni.
The $8^+$ level is isomeric in the former but not in the latter nucleus.
The thickness of the arrows is proportional to the $B$(E2) values
expected on the basis of a seniority classification,
and therefore indicates the likely decay path followed in both nuclei.
In the middle is shown a schematic spectrum of low-energy states with seniority
$\upsilon=0$ (black), $\upsilon=2$ (blue) and  $\upsilon=4$ (red).}
\end{figure}

In the right part of figure 5 is shown the spectrum of $^{72}$Ni
where the $6^+$ level with seniority $\upsilon=4$ is below the $8^+$ level,
opening up an alternative decay path,
which explains why the latter level is no longer isomeric
but has a half-life of less than 20~ns~\cite{Chiara11}.
Similar simple analyses can be done for other semi-magic nuclei
and explain why the $8^+$ level is isomeric in $^{212,214}$Pb and in $^{128}$Pd~\cite{Gottardo12}.

It therefore transpires that the occurrence of isomers in several semi-magic nuclei
is a consequence of the partial conservation of seniority.
The problem can be recast as one of interacting fermions with half-odd-integer spin $j$
and it can be shown~\cite{Qi12} that the partial conservation of seniority
is a very peculiar geometric ({\it i.e.}, interaction-independent) property valid only for $j=9/2$.
A similar analysis can be carried out for particles with integer spin
and it turns out that for bosons partial conservation of seniority
is a prevalent phenomenon~\cite{Isacker},
the consequences of which ({\it e.g.}, for Bose--Einstein condensates of atoms with integer hyperfine spin)
still remain to be investigated.

\section{Emerging Simplicity}
In summary, the concept of dynamical symmetry is arguably too restrictive to analyze systems governed by complex dynamics. From the IBM example presented here, it is tempting to conclude that a satisfactory explanation of the emergence of regularity in complex systems requires the enlarged concepts of partial and quasi dynamical symmetry.

\section*{Acknowledgements}
I wish to thank
Jos\'e-Enrique Garc\'\i a-Ramos,
Stefan Heinze,
Christoph Kremer,
Amiram Leviatan
and Norbert  Pietralla, my collaborators in this work,
and Richard Casten for many helpful suggestions.


\begin{thebibliography}{0}
\expandafter\ifx\csname natexlab\endcsname\relax\def\natexlab#1{#1}\fi
\expandafter\ifx\csname bibnamefont\endcsname\relax
  \def\bibnamefont#1{#1}\fi
\expandafter\ifx\csname bibfnamefont\endcsname\relax
  \def\bibfnamefont#1{#1}\fi
\expandafter\ifx\csname citenamefont\endcsname\relax
  \def\citenamefont#1{#1}\fi
\expandafter\ifx\csname url\endcsname\relax
  \def\url#1{\texttt{#1}}\fi
\expandafter\ifx\csname urlprefix\endcsname\relax\def\urlprefix{URL }\fi
\providecommand{\bibinfo}[2]{#2}
\providecommand{\eprint}[2][]{\url{#2}}

\end{thebibliography}


\begin{references}
\bibitem{Iachello06}
F.~Iachello,
Lie Algebras and Applications (Springer, 2006).

\bibitem{Racah43}
G.~Racah,
Phys.\ Rev.\ 63 (1943) 367;
76 (1949) 1352.

\bibitem{Elliott58}
J.P.~Elliott,
Proc.\ Roy.\ Soc.\ A 245 (1958) 128; 562.

\bibitem{Frank09}
A.~Frank, J.~Jolie and P.~Van~Isacker,
Symmetries in Atomic Nuclei (Springer, 2009).

\bibitem{Leviatan11}
A.~Leviatan,
Prog.\ Part.\ Nucl.\ Phys.\ 66 (2011) 93.

\bibitem{Rochford88}
P.~Rochford and D.J.~Rowe,
Phys.\ Lett.\ B 210 (1988) 5.

\bibitem{Arima75}
A.~Arima and F.~Iachello,
Phys.\ Rev.\ Lett.\ 35 (1975) 1069.

\bibitem{Iachello87}
F.~Iachello and I.~Talmi,
Rev.\ Mod.\ Phys.\ 59 (1987) 339.

\bibitem{Ginocchio80}
J.N.~Ginocchio and M.W.~Kirson,
Phys.\ Rev.\ Lett.\ 44 (1980) 1744;
A.E.L.~Dieperink, O.~Scholten and F.~Iachello,
Phys.\ Rev.\ Lett.\ 44 (1980) 1747.

\bibitem{Arima76}
A.~Arima and F.~Iachello,
Ann.\ Phys.\ (NY) 99 (1976) 253;
111 (1978) 201;
123 (1979) 468.

\bibitem{Warner83}
D.D.~Warner and R.F.~Casten,
Phys.\ Rev.\ C 28 (1983) 1798;
P.O.~Lipas, P.~Toivonen and D.D.~Warner,
Phys.\ Lett.\ B 155 (1985) 295.

\bibitem{Casten81}
R.F.~Casten,
in Interacting Bose-Fermi Systems in Nuclei,
edited by F.~Iachello (Plenum, 1981).

\bibitem{Jolie01}
J.~Jolie {\it et al.},
Phys.\ Rev.\ Lett.\ 87 (2001) 62501;
89 (2002) 182502.

\bibitem{Cejnar98}
P.~Cejnar and J.~Jolie,
Phys.\ Lett.\ B 420 (1998) 241.

\bibitem{Kremer}
C.~Kremer {\it et al.},
to be published.

\bibitem{Leviatan86}
A.~Leviatan, A.~Novoselsky and I.~Talmi,
Phys.\ Lett.\ B 172 (1986) 144.

\bibitem{Pan98}
F.~Pan and J.P.~Draayer,
Nucl.\ Phys.\ A 636 (1998) 156.

\bibitem{Rowe88}
D.J.~Rowe, P.~Rochford and J.~Repka,
J.\ Math.\ Phys.\ (N.Y.) 29 (1988) 572.

\bibitem{Rowe98}
D.J.~Rowe, C.~Bahri and W.~Wijesundera,
Phys.\ Rev.\ Lett.\ 80 (1998) 4394;
C.~Bahri and D.J.~Rowe,
Nucl.\ Phys.\ A 662 (2000) 125;
M.~Macek,  J.~Dobe\v s and P.~Cejnar,
Phys.\ Rev.\ C 80 (2009) 014319;
D.~Bonatsos, E.A.~McCutchan and R.F.~Casten,
Phys.\ Rev.\ Lett.\ 104 (2010) 022502.

\bibitem{Leviatan96}
A.~Leviatan,
Phys.\ Rev.\ Lett.\ 77 (1996) 818;
A.~Leviatan and P.~Van~Isacker,
Phys.\ Rev.\ Lett.\ 89 (2002) 222501.

\bibitem{Shirokov98}
A.M.~Shirokov, N.A.~Smirnova and Yu.F.~Smirnov,
Phys.\ Lett.\ B 434 (1998) 237.

\bibitem{Isacker99}
P.~Van~Isacker,
Phys.\ Rev.\ Lett.\ 83 (1999) 4269;
J.E.~Garc\'\i a-Ramos, A.~Leviatan and P.~Van~Isacker,
Phys.\ Rev.\ Lett.\ 102 (2009) 112502;
A.~Leviatan, J.E.~Garc\'\i a-Ramos and P.~Van~Isacker,
Phys.\ Rev.\ C 87 (2013) 021302 (R).

\bibitem{Shalit63}
A.~de-Shalit and I.~Talmi,
Nuclear Shell Theory (McGraw-Hill, 1963);
I.~Talmi,
Simple Models of Complex Nuclei (Harwood, 1993).

\bibitem{Grawe97}
H.~Grawe {\it et al.},
Prog.\ Part.\ Nucl.\ Phys.\ 38 (1997) 15.

\bibitem{Mills07}
W.J.~Mills {\it et al.},
Phys.\ Rev.\ C 75 (2007) 047302.

\bibitem{Escuderos06}
A.~Escuderos and L.~Zamick,
Phys.\ Rev.\ C 73 (2006) 044302;
P.~Van~Isacker and S.~Heinze,
Phys.\ Rev.\ Lett.\ 100 (2008) 052501.

\bibitem{Chiara11}
C.J.~Chiara {\it et al.},
Phys.\ Rev.\ C 84 (2011) 037304.

\bibitem{Gottardo12}
A.~Gottardo {\it et al.},
Phys.\ Rev.\ Lett.\ 109 (2012) 162502;
H.~Watanabe {\it et al.},
Phys.\ Rev.\ Lett.\ 111 (2013) 152501.

\bibitem{Qi12}
C.~Qi, Z.X.~Xu and R.J.~Liotta,
Nucl.\ Phys.\ A 884 (2012) 21.

\bibitem{Isacker}
P.~Van~Isacker and S.~Heinze,
to be published.
\end{references}
\end{document}